# Formation of Enhanced Uniform Chiral Fields in Symmetric Dimer Nanostructures

**Xiaorui Tian[1], Yurui Fang[2,*], and Mengtao Sun[3,*]**

[1]Division of i-Lab, Suzhou Institute of Nano-Tech & Nano-Bionics, Chinese Academy of Sciences, Suzhou 215123, Jiangsu, China
[2]Department of Applied Physics, Chalmers University of Technology, SE-412 96, Göteborg, Sweden
[3]Beijing National Laboratory for Condensed Matter Physics, Institute of Physics, Chinese Academy of Sciences, Beijing, 100190, Beijing, China
[*]Corresponding authors: yurui.fang@chalmers.se (Y. Fang), mtsun@iphy.ac.cn (M. Sun)

## ABSTRACT

Chiral fields with large optical chirality are very important in chiral molecules analysis, sensing and other measurements. Plasmonic nanostructures have been proposed to realize such super chiral fields for enhancing weak chiral signals. However, most of them cannot provide uniform chiral near-fields close to the structures, which makes these nanostructures not so efficient for applications. Plasmonic helical nanostructures and blocked squares have been proved to provide uniform chiral near-fields, but structure fabrication is a challenge. In this paper, we show that very simple plasmonic dimer structures can provide uniform chiral fields in the gaps with large enhancement of both near electric fields and chiral fields under linearly polarized light illumination with polarization off the dimer axis at dipole resonance. An analytical dipole model is utilized to explain this behavior theoretically. 30 times of volume averaged chiral field enhancement is gotten in the whole gap. Chiral fields with opposite handedness can be obtained simply by changing the polarization to the other side of the dimer axis. It is especially useful in Raman optical activity measurement and chiral sensing of small quantity of chiral molecule.

## Introduction

Optical activity (OA), that the material responses differently to left circularly and right circularly polarized light (LCP and RCP, respectively)[1], is very universal and important in nature. Most amino acids and sugars which are building blocks of life have optical activity due to their chiral structures. Chirality sensitive circular dichroism (CD), optical rotatory dispersion (ORD) and Raman optical activity (ROA) are three significant spectroscopic techniques for chiral material analysis and have been widely used in biomolecule science[1,2]. However, the optical activity of most nature molecules is inherently weak, which limits those applications to samples in a level of large quantity.

Plasmonic chiral nanostructures, which can show optical activity that are several orders of magnitude stronger than nature molecules, offer an important way of realizing small quantity of molecules detection, and attract intense attention in recent years[3,4]. Surface plasmons (SPs), collective oscillations of free electrons confined at metal-dielectric interfaces, are responsible for such strong chiral response. In the past several years, far-field chiroptical response of huge optical activity has been observed and studied in various chiral plasmonic nanostructures, such as two-dimensional or planar nanostructures[5-10] , bi- and multilayered structures[11-17], as well as three-dimensional helical nanostructures of different forms[18-23], and so on[24]. Plasmon-enhanced chiroptical response of nearby chiral molecules has also been studied both theoretically and experimentally[25-30].

In addition to the far-field, chiroptical near-fields enhancement as well as their handedness control is also, even more important, especially for enhancing weak chiral effects of molecules in applications of molecular analysis, because the chirality enhancement of adsorbed chiral molecules on the nanostructures is an averaged effect of the chiral field close to the whole structure. The approximate absorption of a chiral molecule in a chiral electromagnetic field is[31,32]

$$A^{\pm} = \frac{\omega}{2}\left(\alpha''\left|\widehat{\boldsymbol{E}}\right|^2 + \gamma''\left|\widehat{\boldsymbol{B}}\right|^2\right) \pm G''\omega Im\left(\vec{\boldsymbol{E}}^{*} \cdot \vec{\boldsymbol{B}}\right) \qquad (1)$$

where $\alpha''$, $\gamma''$ and $G''$ are the imaginary parts of electric, magnetic and mixed electric-magnetic dipole polarizability of the molecules, respectively. The last item in equation (1) determines the different absorption (circular dichroism, CD) of molecules to chiral fields with different handedness. For a given molecule, it is proportional to the so-called optical chirality C of the local electromagnetic field, which is defined as[33,34]

$$C = -\frac{\varepsilon_0\omega}{2} \cdot Im\left(\vec{\boldsymbol{E}}^{*} \cdot \vec{\boldsymbol{B}}\right) \qquad (2)$$

This definition applies to any monochromatic field, no matter with or without nanostructures, and no matter chiral or achiral structures. For plane waves of circularly polarized light (CPL), the optical chirality $C_{CPL} = \pm\frac{\varepsilon_0\omega}{2c} \cdot |\vec{\boldsymbol{E}}|^2$ (+ for LCP and – for RCP, respectively). The basic information obtained from equation (2) is that for fields with non-zero optical chirality, there should be parallel components of electric and magnetic fields. Enhanced chiral fields with optical chirality larger than $C_{CPL}$, i.e. so-called chiral "hot" spots,  have been obtained in different plasmonic chiral structures[25,35-38], even in a single achiral nanoparticle[39]. However, for most of them (both chiral and achiral) there are chiral hot spots with both left and right handedness around the same structure, which will decrease or even cancel the enhancement because molecules are generally distributed randomly over the whole structure. This makes them not good candidates for chirality enhancement. In 2012, Schäferling proposed a very simple square structure to yield chiral near-fields for sensing with linear polarization[40], and in 2014, they theoretically provide another helical structure to yield ultra-uniform chiral near-field inside the helix with large optical chirality which seems to be an ideal system for molecule analysis and sensing application[41]. But fabrication of the sample is a challenge. In this work, we show that simple Au block dimer structures can provide very uniform chiral fields with large optical chirality in the gaps under linearly polarized light illumination with polarization off the dimer axis. Chiral fields with opposite handedness can be obtained simply by changing the polarization to the other side of the dimer axis. An analytical dipole model is suggested to explain this behavior theoretically. Polarization, thickness, length and gap dependent situations are investigated. The results show plasmonic block dimer structures may have promising applications in related fields such as environment sensors, especially in ROA measurement.

## Results

### Chiral fields generated by one dipole and a dipole dimer under external excitation

Figure 1 illustrates the formation principle of strong chiral fields in the gap of a dipole dimer. Schäferling has analytically shown that, for a Hertzian dipole driven by a linearly polarized external field, the chiral near-field around the dipole is a four lobes with alternating sign of optical chirality around the scatter, here shown in Figure 1a-i, due to the interaction of the incident magnetic field and the scattered electric field by the dipole[40]. We will show that when two  such dipoles are put together with the same oscillation direction, if we choose the excitation direction so as that the two lobs with the same chirality overlapped, an enhanced chiral field with the same and uniform chirality will be formed in the gap between the dipoles, as shown in Figures 1a-ii to –iv. To better understand the interaction and interference of the two dipoles, we first analyze the model of two coupled dipoles with coupled-dipole approximation method (see Supporting Information for details)[42,43]. The dipole moments of the two coupled dipoles can be expressed as[44]

$$\boldsymbol{p}_{e,1} = \overleftrightarrow{\boldsymbol{\alpha}_1}\left(\boldsymbol{E}_{1,in} + \frac{k^2}{\varepsilon_0}\overleftrightarrow{\boldsymbol{G}_e}(\boldsymbol{r}_1, \boldsymbol{r}_2)\boldsymbol{p}_{e,2}\right) \quad (3)$$

$$\boldsymbol{p}_{e,2} = \overleftrightarrow{\boldsymbol{\alpha_2}}\left(\boldsymbol{E}_{2,in} + \frac{k^2}{\varepsilon_0}\overleftrightarrow{\boldsymbol{G}_e}(\boldsymbol{r}_2, \boldsymbol{r}_1)\boldsymbol{p}_{e,1}\right) \quad (4)$$

Where $\overleftrightarrow{\boldsymbol{\alpha_j}}$ is the polarizability tensor, and $\overleftrightarrow{\boldsymbol{G}_e}(\boldsymbol{r}_j, \boldsymbol{r}_k)$ is the electric dyadic Green's function. Solving this set of two equations, we can obtain the self-consistent form of dipole moments

$$\boldsymbol{p}_{e,1} = \frac{\overleftrightarrow{\boldsymbol{\alpha_1}}\boldsymbol{E}_{1,in} + \frac{k^2}{\varepsilon_0}\overleftrightarrow{\boldsymbol{\alpha_1}}\overleftrightarrow{\boldsymbol{G}_e}(\boldsymbol{r}_1, \boldsymbol{r}_2)\overleftrightarrow{\boldsymbol{\alpha_2}}\boldsymbol{E}_{2,in}}{\overleftrightarrow{\boldsymbol{I}} - \frac{k^4}{\varepsilon_0^2}\overleftrightarrow{\boldsymbol{\alpha_1}}\overleftrightarrow{\boldsymbol{G}_e}(\boldsymbol{r}_1, \boldsymbol{r}_2)\overleftrightarrow{\boldsymbol{\alpha_2}}\overleftrightarrow{\boldsymbol{G}_e}(\boldsymbol{r}_2, \boldsymbol{r}_1)} \quad (5)$$

$$\boldsymbol{p}_{e,2} = \frac{\overleftrightarrow{\boldsymbol{\alpha_2}}\boldsymbol{E}_{2,in} + \frac{k^2}{\varepsilon_0}\overleftrightarrow{\boldsymbol{\alpha_2}}\overleftrightarrow{\boldsymbol{G}_e}(\boldsymbol{r}_2, \boldsymbol{r}_1)\overleftrightarrow{\boldsymbol{\alpha_1}}\boldsymbol{E}_{1,in}}{\overleftrightarrow{\boldsymbol{I}} - \frac{k^4}{\varepsilon_0^2}\overleftrightarrow{\boldsymbol{\alpha_2}}\overleftrightarrow{\boldsymbol{G}_e}(\boldsymbol{r}_2, \boldsymbol{r}_1)\overleftrightarrow{\boldsymbol{\alpha_1}}\overleftrightarrow{\boldsymbol{G}_e}(\boldsymbol{r}_1, \boldsymbol{r}_2)} \quad (6)$$

Where $\overleftrightarrow{\boldsymbol{I}}$ is the unit dyad. In our case, the two same dipoles are located in the same plane perpendicular to the incident wave vector **k**, so the incident fields are the same. The total generated electric field by the two dipoles at position **r** is

$$\boldsymbol{E}_d(\boldsymbol{r}) = \boldsymbol{E}_1(\boldsymbol{r}) + \boldsymbol{E}_2(\boldsymbol{r}) = \frac{k^2}{\varepsilon_0}\left(\overleftrightarrow{\boldsymbol{G}_e}(\boldsymbol{r}, \boldsymbol{r}_0)\boldsymbol{p}_{e,1} + \overleftrightarrow{\boldsymbol{G}_e}(\boldsymbol{r}, \boldsymbol{r}_0)\boldsymbol{p}_{e,2}\right) \quad \textbf{(7)}$$

The chirality of the field is then

$$C(\boldsymbol{r}) = -\frac{\varepsilon_0\omega}{2} \cdot Im\left((\boldsymbol{E}_{in} + \boldsymbol{E}_d)^* \cdot (\boldsymbol{B}_{in} + \boldsymbol{B}_d)\right) = -\frac{\varepsilon_0\omega}{2} \cdot Im(\boldsymbol{E}_d{}^* \cdot \boldsymbol{B}_{in}) \quad (8)$$

Term $\boldsymbol{E}_d{}^* \cdot \boldsymbol{B}_d = 0$ as the electric and magnetic field vectors are orthogonal. $\boldsymbol{E}_{in}{}^* \cdot \boldsymbol{B}_d = 0$ because $\boldsymbol{B}_d$ is always orthogonal to the dipole $\boldsymbol{p}_e$ and thus has not parallel component with $\boldsymbol{E}_{in}$. Here we choose linearly polarized light with the polarization off the dipole dimer axis by 45° as the incident field. Both dipole of plasmonic sperical nanoparticles and pure dipoles can be analyzed with this analytic model (see Supporting Information). The case of pure point dipoles is given in Figure 1a, which shows how two dipoles coupled together and form a strong chiral field in the middle of the gap. Positive (negative) parts in each picture means the fields are left (right)-handed. Smaller gap will cause stronger chiral field enhancement (Figure 1a-ii, 1a-iii). A simple analysis shows that the chiral field in the middle of the two dipoles has a $\frac{p(d)}{d^3}$ dependence on the gap distance d (see Supporting Information). Opposite handed chiral field in the gap can be obtained by just rotating the incident polarization by 90˚ (Figure 1a-iii and -iv).

Following this, small Au nanospheres with radius of 5 nm as plasmonic dipoles are numerically studied using full wave simulations. As shown in Figure 1b-i, chiral fields with alternating handedness (four lobes with alternating sign of optical chirality) around the plasmonic dipole are formed under linear polarization illumination, which is similar to the above dipole result and what has been observed by Schäferling[40]. The underlying physical mechanism is similar to the dipole case, that the scattered field is distorted by the scatter, while the incident field keeps unchanged, so there are parallel electric and magnetic components; additionally, the plasmonic dipole has a delayed phase response to the incident fields. The two aspects together result in a non-zero chiral field, of which the optical chirality reaches a maximum at the resonant wavelength where the phase difference between the scattered field and the incident field is around pi/2 and, the local electric field is mostly enhanced. Actually, within the whole range of plasmon decay wavelength, the phase difference is between 0 and pi, thus non-zero chiral fields also exist away from the resonant wavelength position. Moreover than the Hertzian dipole condition, the incident electric field and scattered magnetic field $Im(\vec{\boldsymbol{E}}_{in}^{*} \cdot \vec{\boldsymbol{B}}_{\boldsymbol{scat}})$ also give a non-zero but very small chiral field value for this plasmonic scatter because the field is distorted by the particle. When two spheres are put together, and under the illumination of linearly polarized light with the polarization off the axis of the two dipoles by 45˚, the overlapped chiral fields between the dipoles have the same optical chirality sign, as shown in Figure 1b-ii; when the gap decreases, the chiral near-field becomes stronger (Figure 1b-iii). Besides, due to the strong interaction of the two dipoles, intense local field in the gap is obtained (Figure 1c). As the electric field is always perpendicular to the metal surface, the electric field in the dimer gap is seriously distorted by the dimer and no longer in the incident polarization direction, while the incident magnetic field is always perpendicular to the polarization direction, the dot product $Im(\vec{\boldsymbol{E}}_{scat}^{*} \cdot \vec{\boldsymbol{B}}_{in})$ is non-zero, resulting in strong chiral fields. Figure 1d schematically shows the directions of scattered and incident fields. The sign of optical chirality of the chiral field and thus the handedness of the chiral field can be changed to the opposite by rotating the incident polarization to the opposite side of the dimer axis (Figure 1b-iii and -iv), the same as the coupled dipoles condition, which is because the angle between the electric and magnetic field vectors changes from acute angle to obtuse angle, or conversely. All of the results shown in Figure 1b match that in Figure 1a very well, which demonstrates that our analysis works very well.

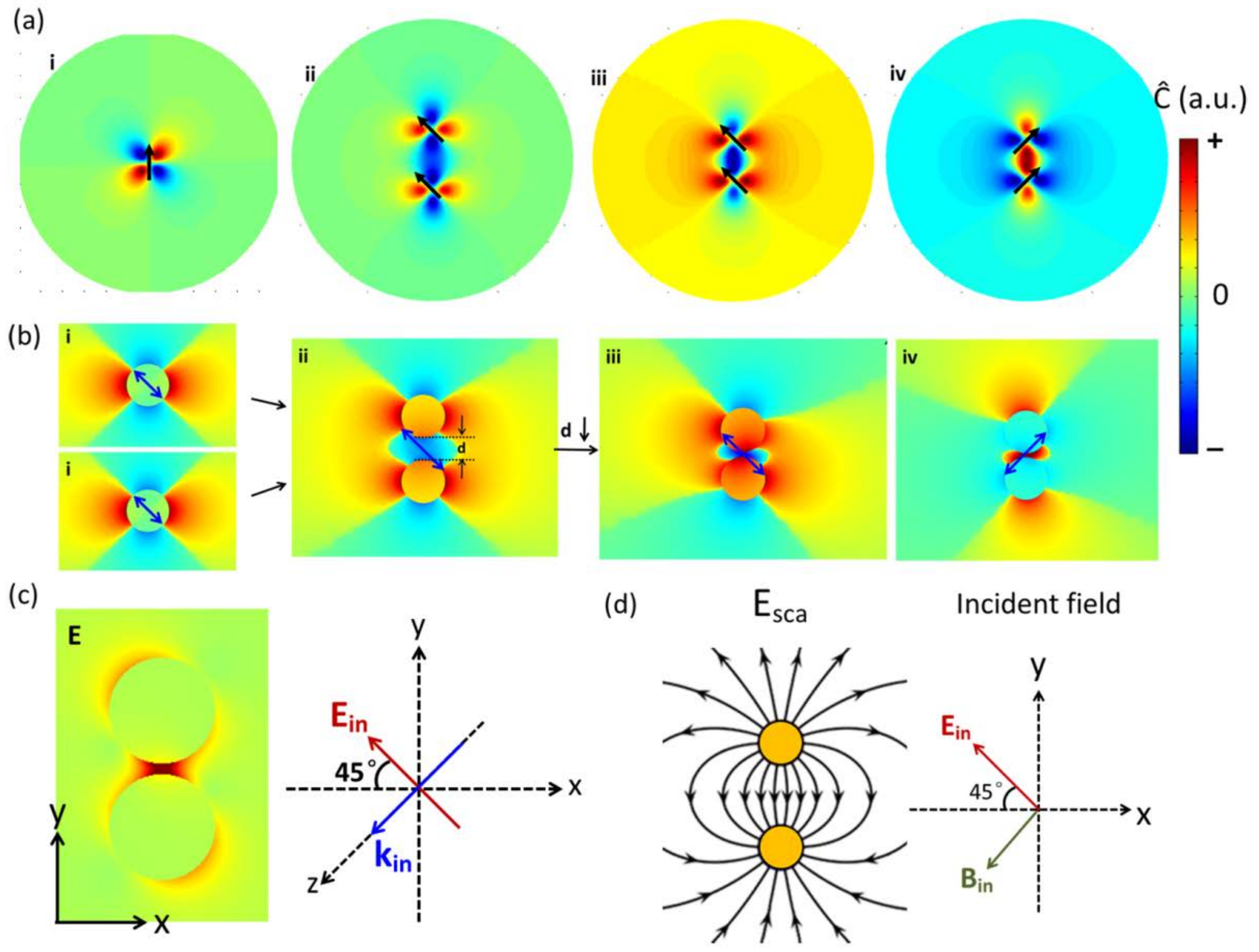

**Figure 1.** Formation schematic of enhanced chiral near-fields with uniform optical chirality in the gap of a coupled point dipole dimer (a) and Au spherical nanoparticle (10 nm diameter) dimer (b). (a) Analytically calculated chiral near-fields distributions of (i) one dipole, (ii) two dipoles with a large gap d of 0.06λ, and (iii-iv) two dipoles with a small gap d of 0.04λ. Black arrows show the dipole momentum. Signs of '+' and '–' in the scale bar indicate the field is left- and right- handed, respectively, which applies to all figures in the following. (b) Numerically calculated chiral near-fields distributions of (i) one sphere, (ii) two spheres with a large gap d of 10 nm, and (iii-iv) two dipoles with a small gap d of 2 nm. (c) Corresponding electric fields distribution of the case (b)-iii. The right coordinates gives incident polarization and direction. (d) Schematic of the directions of incident fields and scattered fields by a dipole dimer.

## Chiral fields formed by dimers of different shapes

The principle discussed above works quite well with many kinds of plasmonic dimers of different shapes. Figure 2 shows the results of Au sphere dimer, disk dimer and block dimer. These three dimers have the same gap distance of 5 nm and similar size. Electromagnetic enhancement and optical chirality of the field in the gap are studied. Optical chirality enhancement $\hat{C} = C/|C_{CPL}|$ is used to characterize the chiral field.

When two same nanoparticles are coupled together, only the bright bonding mode can be excited by a plane wave, manifested as two peaks (longitudinal and transverse modes) in the extinction spectra, as shown in Figure 2d to f. From the electric field distributions in dimer gaps shown in the upper row in Figure 2a to c, one can see strong electromagnetic enhancement can be obtained in each dimer, with disk dimer showing the strongest enhancement but in a small volume, while block dimer giving the largest enhancement volume. As to the optical chirality of the field which is our concern in this paper, strongly enhanced chiral near-field is observed for all dimers, especially in the gaps, which can be seen from the lower two rows in Figure 2a to c. Comparing the three cases, chiral fields in the sphere and disk dimer gaps are quite confined in a small volume close to the shortest distance, while block dimer gives both large and uniform optical chirality distribution due to large interacting area with constant gap distance, which is very useful for chiral molecule sensing and especially, for the ROA measurement. Another advantage of the dimer structures proposed here is that the handedness of the chiral field can be changed to the opposite simply by rotating the incident polarization. For example, in a typical ROA measurement, one just measures the spectra under linearly polarized light with two symmetrically crossed polarizations off the dimer axis. Though the previous reported helical structure gives uniform and strong chiral fields, the handedness of the field cannot be changed once the structures is fixed, which limits its application. As discussed in Figure 1, uniform chiral fields can exist in quite a broad dipole resonance wavelength range (data not shown). In the following, we will focus on the block structure since it shows better property of strong and uniform chiral fields in a large volume.

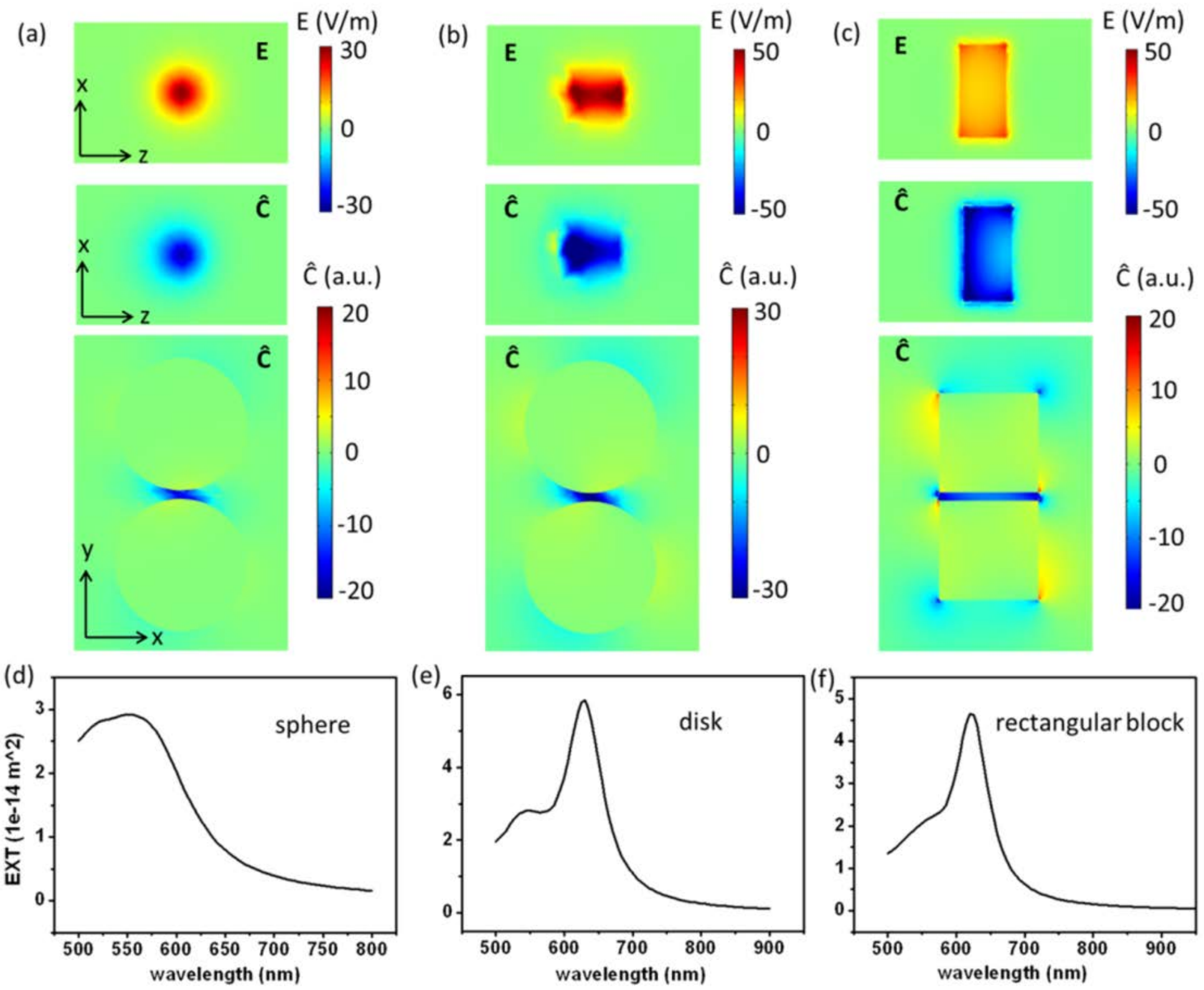

**Figure 2.** Chiral fields formed by different shaped dimers: (a) 40 nm radius Au spheres, (b) 40 nm radius, 30 nm height Au disks and (c) 60 nm (length) × 60 nm (width) × 30 nm (height) Au blocks dimer. The uppermost images in (a)-(c) show the electric near-field distributions of dimers. The lower two rows of images in (a)-(c) show the chiral field enhancement distributions of different cut planes. (d)-(f) Corresponding extinction spectra. The x-y slices are cut from the middle position of height.

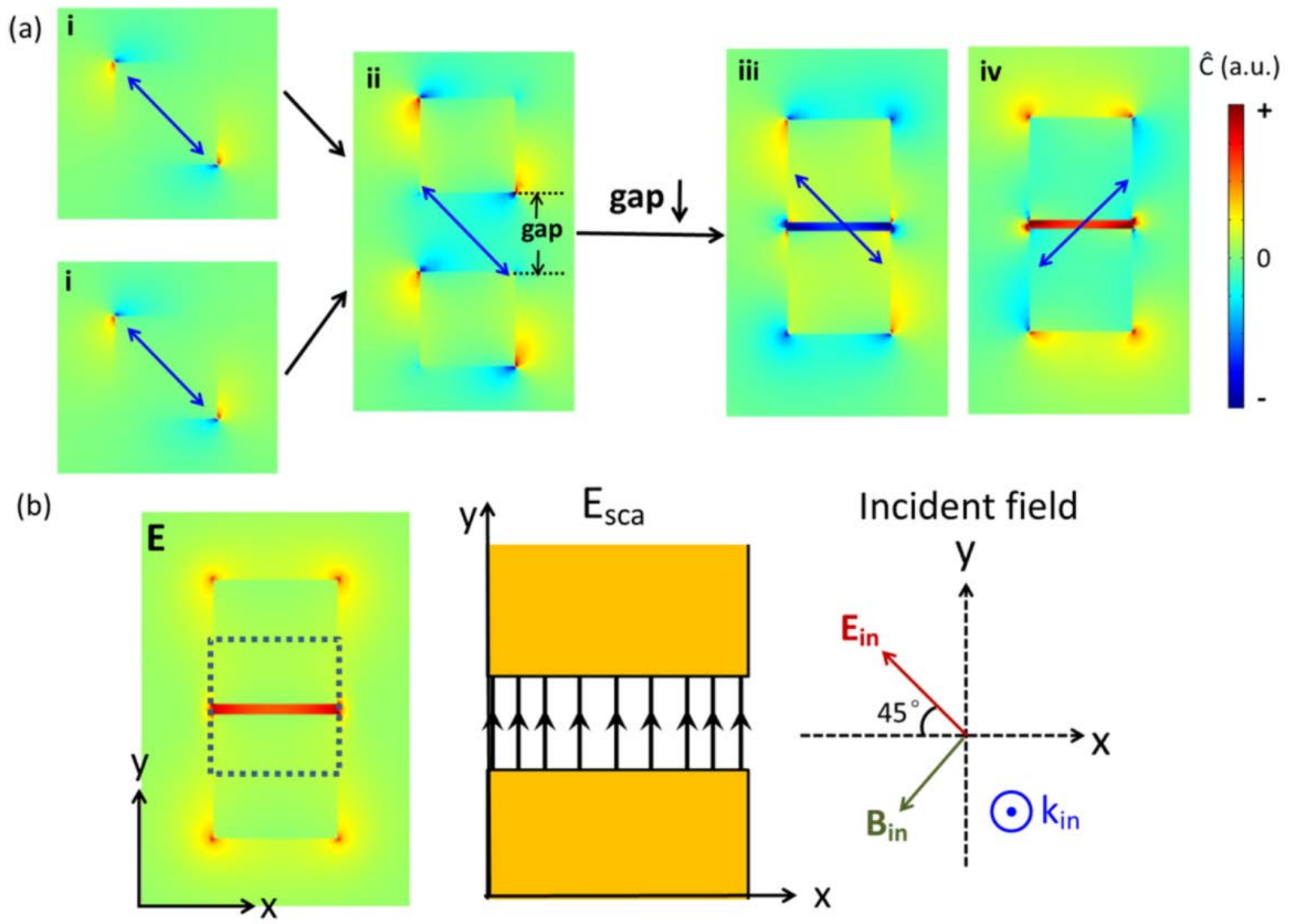

**Figure 3.** Formation schematic of enhanced chiral near-fields with uniform optical chirality in the gap of a Au block (60 nm × 60 nm × 30 nm) dimer located on glass substrate in water surroundings, excited under polarizations indicated by blue arrows. (a) Chiral near-fields distributions of (i) one Au block, (ii) two blocks with a large gap d of 50 nm, and (iii-iv) block dimers with a small gap d of 5 nm. (b) Corresponding electric fields distribution of the case (a)-iii (left image); the right schematics show directions of incident fields and scattered fields (only fields in the gap are concerned) by a block dimer. All slices are cut from middle positions of the height.

## Polarization-dependent optical chirality of block dimer

Considering most experimental situations, block dimers on glass substrate embedded in water are studied in the following, instead of the ideal case of in air. Figure 3 shows how chiral fields evolve from a symmetric distribution around a single block to a confined and with uniform optical chirality in the gap of a block dimer. It can be seen that the dipole model given in Figure 1 well tells the formation mechanism of the uniform chiral fields in dimer structures.

In order to get deeper understanding of this system, polarization-dependent chiral fields of the block dimer are investigated, shown in Figure 4. Only fields in dimer gaps are considered, because strongest chiral fields are confined here. The extinction spectra show the variation trend of the longitudinal and transverse modes of the dimer under different polarizations. To well mimic the real experiment, volume averaged optical chirality defined as $\langle \hat{C} \rangle = \frac{1}{V}\int_V \ \hat{C} \cdot dV$ is also considered in gaps. Unsurprisingly, 0˚ or 90˚ excitation gives near zero chiral field value (shown in Figure 4b-i, b-v, c-i and c-v). This is because at 90˚ the scattered electric field has the same oscillating direction with the incident polarization, resulting in always orthogonal electric and magnetic field components, and at 0˚ the scattered electric field in the gap is near zero. When polarization of the incident field is off the two symmetry axes (x and y axis), the excited bonding mode and orthogonal incident field component offer parallel electric and magnetic components with delayed phase, which results in non-zero optical chirality, as shown in Figure 4 b-ii to iv, c-ii to iv. It is easy to understand that 45˚ excitation gives the strongest chiral field. The volume averaged chirality enhancement reaches 18 times for 45˚, which is a very big value.

The case of RCP excitation is also studied for a comparison, shown in Figure 4b-iii. As reported in previous papers, under circularly polarized light excitation, dimers can provide enhanced CD of molecules[45]. The above formation mechanism given in Figure 1 and Figure 3 can also be used to understand the chirality enhancement under circularly polarized light excitation, but with time varying scattered and incident field directions. The dot product $Im(\vec{\boldsymbol{E}}^{*}_{scat} \cdot \vec{\boldsymbol{B}}_{\boldsymbol{in}})$ is a time-averaged value, resulting in a smaller value. Uniform chiral fields in the gap can be obtained as well, but at some other different wavelengths off resonance, depending on the phase delay between the incident field and the excited near-fields, which is different for different dimers. For this dimer structure, relative uniform chiral filed occurs at 835 nm. At the resonant wavelength of 750 nm, the phase delay and the rotation of the incident light results in a non-uniform chiral field. The volume averaged optical chirality at the resonance is very small as well, as shown by the red dashed curve in Figure 4b-iii. The uniform chiral field at off resonant wavelength is not convenient to be used in practice as it is hard to decide the wavelength.

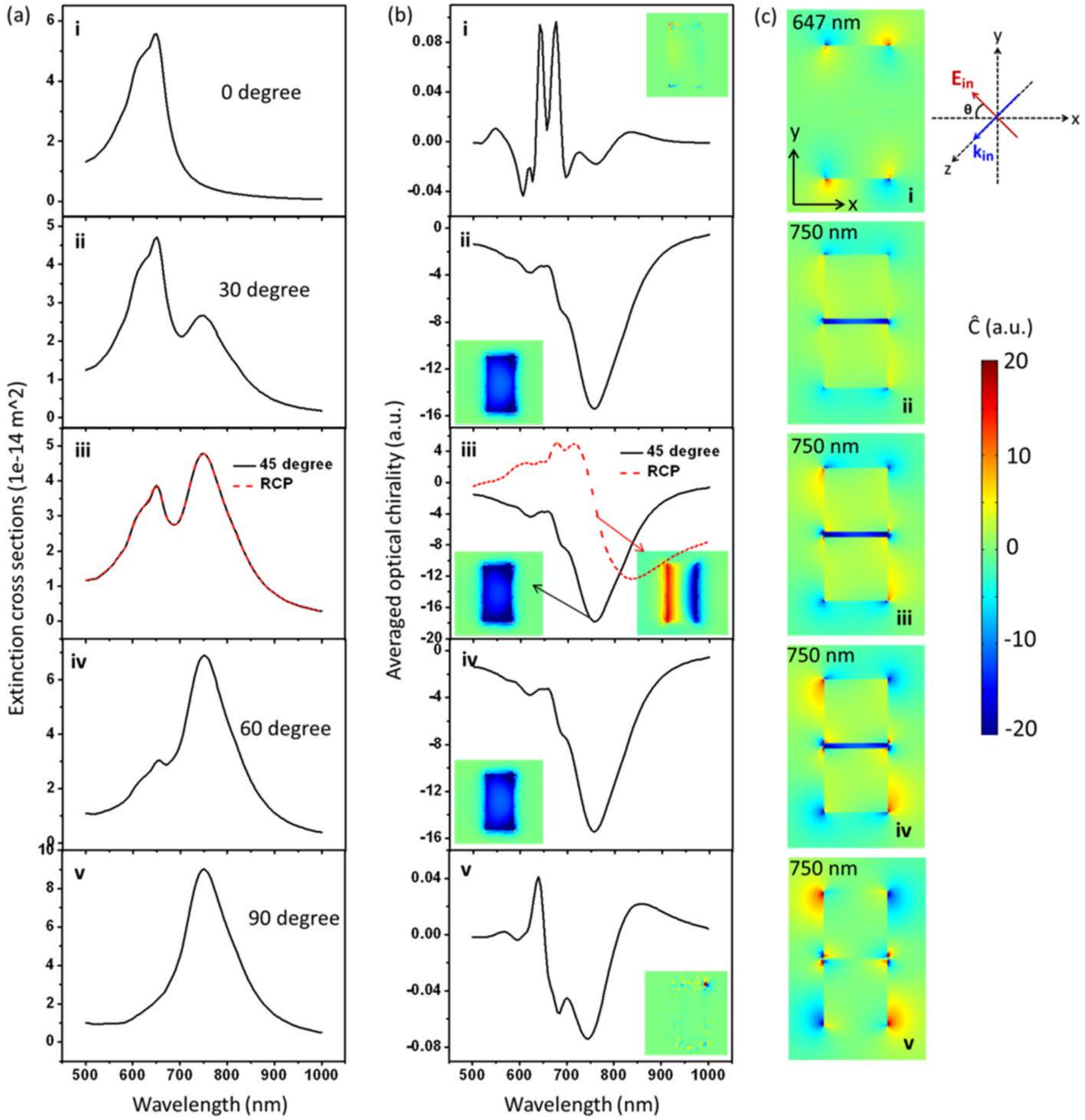

**Figure 4.** Polarization-dependent optical chirality in the gap of Au block dimers (60 nm × 60 nm × 30 nm, gap d = 5 nm) on glass in water environment. (a) Extinction spectra. (b) Volume averaged optical chirality in the gap. Insets show chiral near-field distributions in a plane parallel to the gap at the resonant peak positions, cut from the middle position of the gap. Red curves are for the case of RCP excitation. (c) Corresponding optical chiral near-field distributions in x-y

plane at the dipole resonant wavelength, cut from middle of the height. The scale bar applies to all images except the inset images in b-i and b-v, whose intensity is magnified by 5 times to get a better view.

## Thickness-dependent optical chirality of block dimer

The thickness of block has a significant effect on the chiral field. Block dimers with thickness of 15 nm, 30 nm and 60 nm are studied, as shown in Figure 5. Both length and width of each block are 60 nm. The gap distance is 5 nm. From the volume averaged optical chirality shown in Figure 5b and optical chirality distributions in Figure 5c, it all can be seen that thinner block shows much larger optical chirality, mainly because the scattered filed can be well confined in the gap for thinner blocks. When the thickness increases, the induced field in the deeper position becomes weaker because of the screen effect of the metal, resulting in weak even zero chiral field distribution. Moreover, when the thickness becomes thicker and thicker, the delay effect is more and more obvious, and when the dimer is on a substrate, hybrid new resonant modes will appear because of the coupling between the substrate and dimer. The RCP excitation situations are shown for a comparison. One can see that for thinner dimers, the chiral field enhancement with linearly polarized light is always larger than the RCP, and the RCP conditions have non-uniform chiral near-fields in the gap (Figure 5d). As the thickness increases, the enhancements become similar.

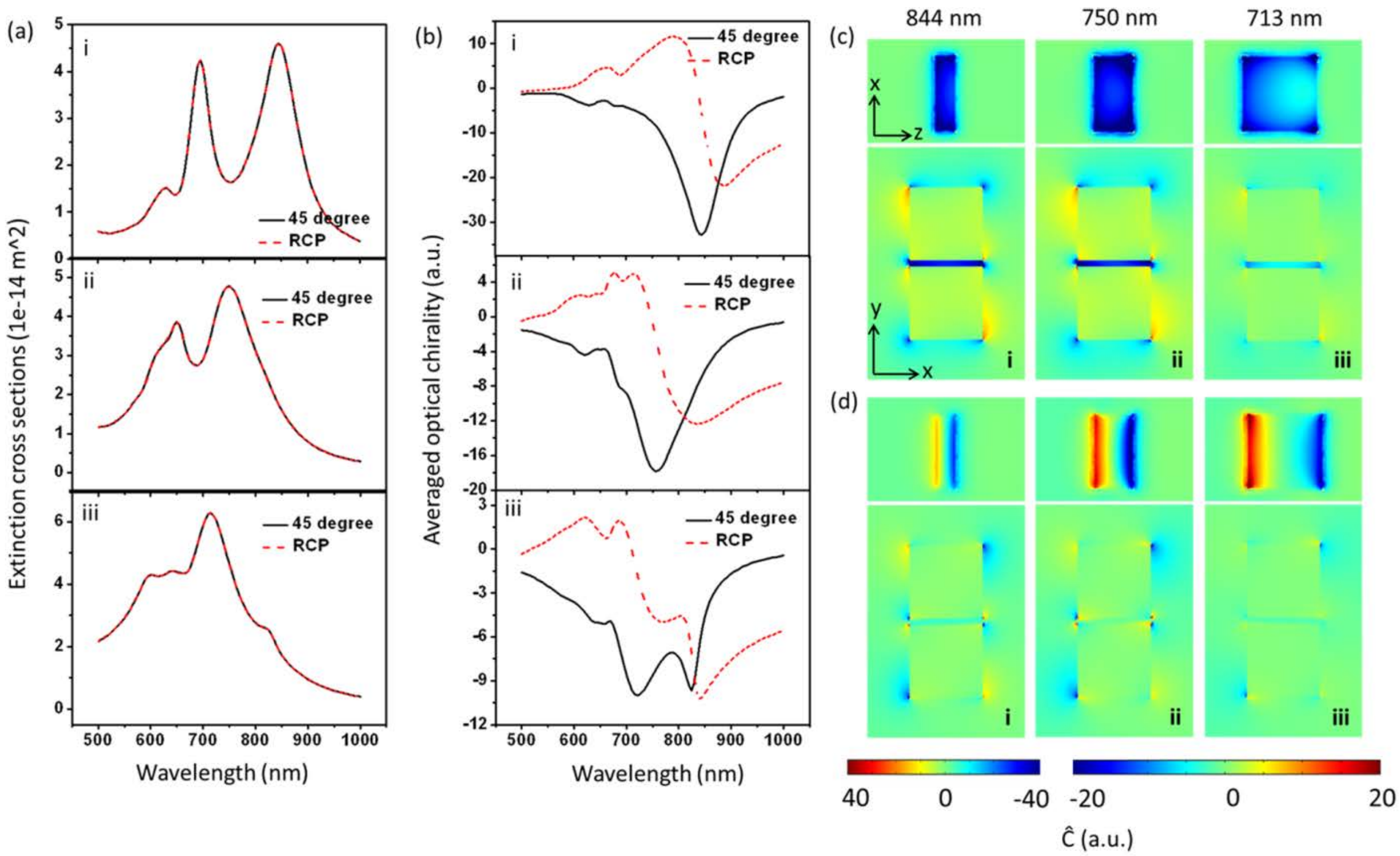

**Figure 5.** Thickness-dependent optical chirality in the gap of Au block-dimers (60 nm × 60 nm × T nm, gap d = 5 nm) on glass in water environment. (a) Extinction spectra. (b) Volume averaged optical chirality in the gap. Red curves are for the case of RCP excitation. (c and d) Super chiral near-field distributions at the dipole resonant wavelength excited by linear polarized light (c) and RCP light (d). Rows (in (a) and (b)) or columns (in (c) and (d)) of i, ii and iii correspond to T = 15 nm, T = 30 nm and T = 60 nm respectively. The x-y slices are cut from middle of the height; x-z slices are cut from the middle position of the gap.

### Length-dependent optical chirality of block dimer

Figure 6 shows the effect of block length on the optical chirality of the fields in the gap. From Figure 6a it can be seen that when the length increases, extinction cross section becomes larger, which means the dipole momentum becomes stronger. Figure 6b and c show that, when the length increases, the enhancement of the optical chirality also increases, and the chiral field becomes more uniform. In Fig. 6c-i, the chiral field is not so uniform because the length (defined in y axis) of the particle is less than the width, which causes the coupled dipoles mainly oscillating in x direction even if the excitation is in 45° direction. In such an oscillation situation, the chiral field with both '+' and '-' signs exist on the longer sides of the block for single block. When bring such two blocks together, chiral fields with opposite signs will overlap. As the two dipoles are parallel standing side by side, the middle of the block will be weak, and the side by side bonding mods will cause dramatically change of the field in the gap. RCP excitation in comparison shows non-uniform chiral field as well (Figure 6d). For the case of longer length, the volume averaged chiral field enhancement reaches 30, which is as strong as the helical structure[41].

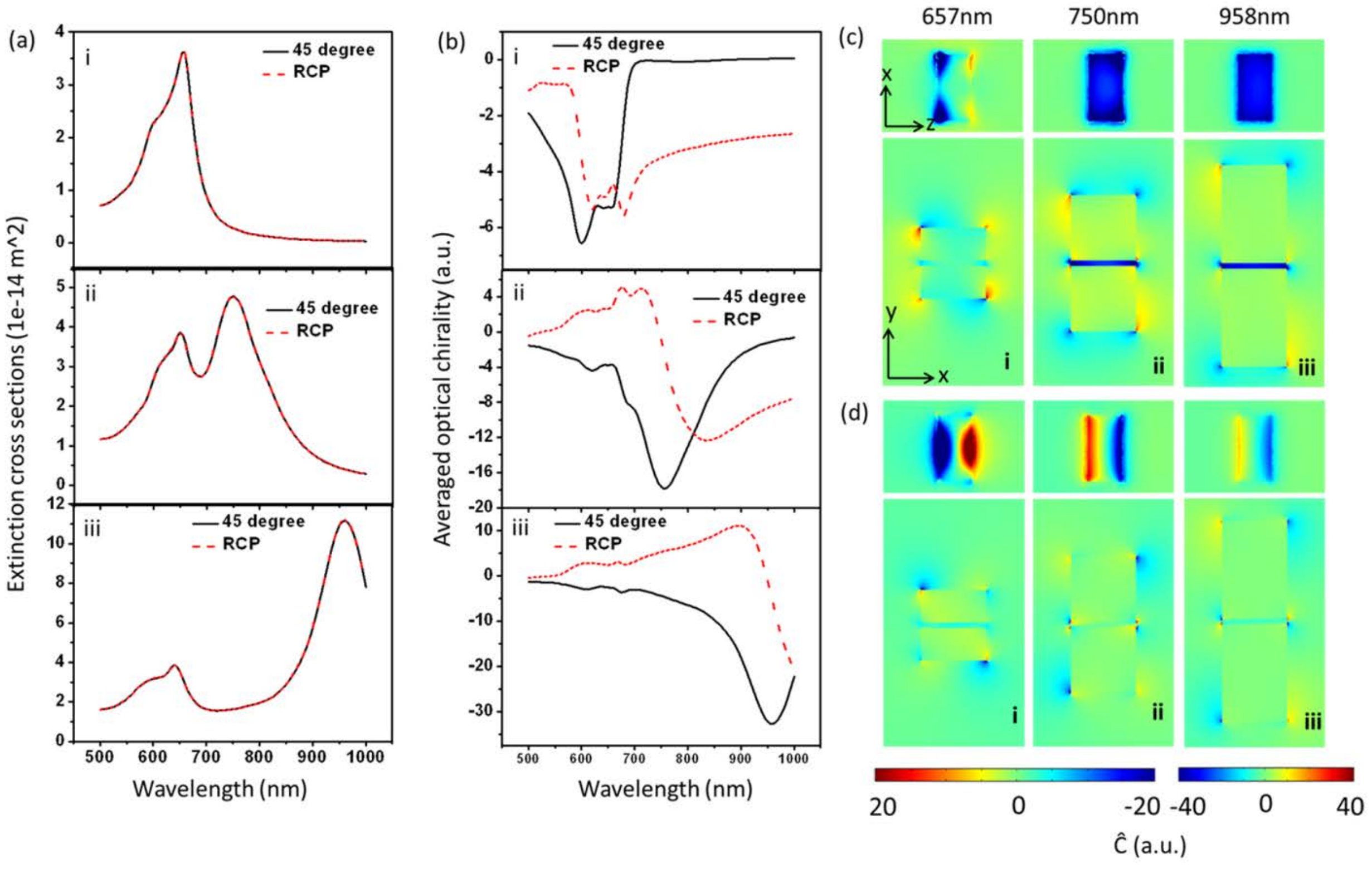

**Figure 6.** Length-dependent optical chirality in the gap of Au block-dimers (L × 60 nm × 30 nm, gap d = 5 nm) on glass in water environment. (a) Extinction spectra. (b) Volume averaged optical chirality in the gap. Red curves are for the case of RCP excitation. (c and d) Super chiral near-

field distributions at the dipole resonant wavelength excited by linearly polarized light (c) and RCP light (d). Rows (in (a) and (b)) or columns (in (c) and (d)) of i, ii and iii correspond to L = 30 nm, L = 60 nm and L = 90 nm respectively. The x-y slices are cut from middle of the height; x-z slices are cut from the middle position of the gap.

## Gap-dependent optical chirality of block dimer

It is in expectation that as the gap decreases, the chiral field becomes stronger and stronger because the field enhancement becomes larger and larger nonlinearly (Figure 7). However, when the gap decreases, it becomes harder for the incident light to induce field deep in the gap because of the screen effect of the metal, resulting in the scattered field weak in the middle and strong in the sharp edges, which make the chiral field is less uniform and weaker in the middle (Figure 7ci). Although the chiral field is non-uniform, it still has the same handedness in the whole gap area, so smaller gap is still a better choice in experiment. The volume averaged chiral field enhancement reaches 30 for the 2 nm gap dimer as well (Figure 7b-i). For all of the gaps, the volume averaged chiral fields have larger values with linearly polarized light excitation than the RCP excitation.

From the ‘uniform’ point of view, one can see that when the gap distance is 5 nm, the chiral field is almost uniform and enhancement is strong enough (Figure 4). In practice, the sizes of chiral molecules rang from below 1 nm to tens of nanometers or even more. But for most of biomolecules, e.g. sugars, amino acides and nucleotides, the sizes are ~0.5 – 1 nm; globular proteins are ~ 2 -- 10 nm. In this range, the gaps we simulated are mostly applicable in sensing and other applications. Even if the gap is fixed to 5 nm, it is still suitable for smaller molecules sensing.

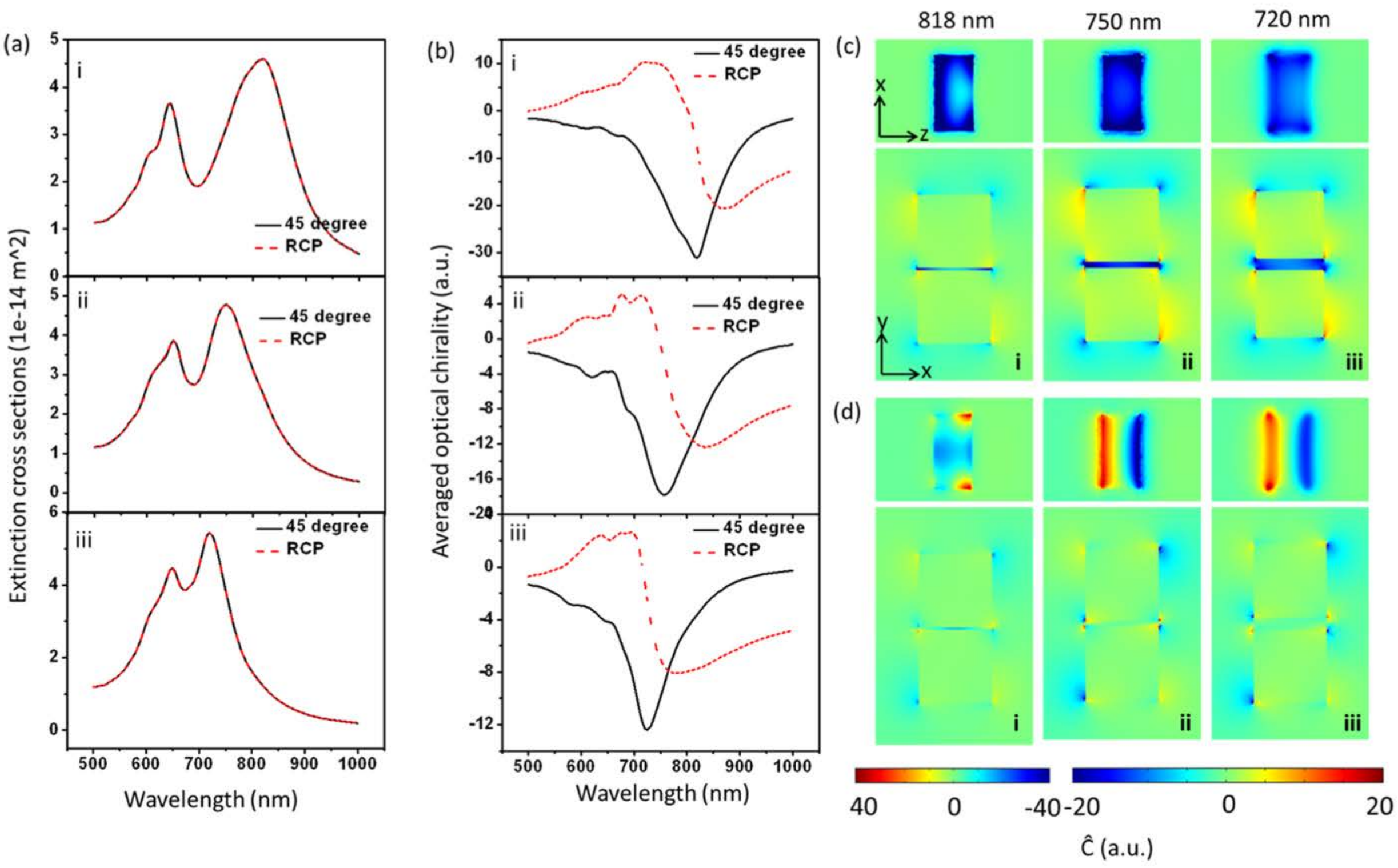

**Figure 7.** Gap-dependent optical chirality in the gap of Au block dimers (60 nm × 60 nm × 30 nm) on glass in water environment. (a) Extinction spectra. (b) Volume averaged optical chirality in the gap. Red curves are for the case of RCP excitation. (c and d) Super chiral near-field distributions at the dipole resonant wavelength excited by linear polarized light (c) and RCP light (d). Rows (in (a) and (b)) or columns (in (c) and (d)) of i, ii and iii correspond to Gap d = 2 nm, d = 5 nm and d = 10 nm respectively. The x-y slices are cut from middle of the height; x-z slices are cut from the middle position of the gap.

## Conclusion and Discussion

In this paper, we proved that very simple plasmonic dimer structures can provide very strong chiral fields with large enhancement on both near electric fields and chiral fields under linear polarization illumination off the dimer axis at dipole resonance. Chiral fields with opposite handedness can be obtained simply by changing the polarization to the other side of the dimer axis. A uniform gap distance of block dimer gives ultra-uniform chiral fields. Polarization, thickness, length and gap dependent optical chirality studies of Au block dimers were investigated in detail to optimize the uniform chiral fields. Situations of dimer structures with different parameters excited by RCP light were also studied in detail, as shown in Figures 5 to 7. All results show that they give weaker, non-uniform chiral fields and small volume averaged optical chirality in the gap at the resonant wavelength. As discussed in the main text, for the RCP condition, when we plot the near chiral field in the longer wavelengths, which is far from the resonant wavelength, but not in the longest peak tail position, they sometimes show uniformity in the gap. However, it is hard to decide at which wavelength they are uniform. There is neither any feature shown in the far field spectra (extinction or scattering) nor other methods by which one can decide the wavelength. Therefore, it is not useful in practice. Under linearly polarized light excitation, when the wavelength shifts from the blue side far from the bonding dipole mode, the chiral field is not uniform anymore because the anti-bonding mode and higher-order modes will cause the field dramatically changing in the gap.

From the chiral near field pictures shown above, one can see that the corners/edges of the blocks are very sharp, which will yield drastic field enhancement in vicinity and then much stronger chiral field. Even for the case of the most uniform field obtained in this paper, the chiral fields at the corners/edges are obviously stronger. In simulations, the mesh size indeed will (if it is not smaller far away than the object features) seriously affect the near field distributions. However, the effect of mesh design in our structure is not so obvious because, compared with the sharp edge or corners, the interfaces are more important in such numerical simulation The sharp features don’t change the handedness of the chiral field, just show stronger enhancement and make the chiral field not very uniform around them. But this effect is positive for actual applications.

The generation of chiral near fields reported so far mainly has three kinds. The first condition is chiral metal nanostructures under circularly polarized light (CPL) illumination; the second condition is non-chiral structures under CPL illumination;  the third one is special structures under linearly polarized light illumination. For the first two situations, the chiral near field is usually non-uniform, and in a lot of cases, there are both left- and right- handed field coexisting in the hot spot (where it is useful)[35,38,39,46]. For the averaged enhancement factor spectra of the chiral near field excited with CPL in reported results, mostly the enhancement varys between positive and negative values in different wavelength[38,46]. While for the linearly polarized light exited helical structures, the chiral field is very uniform and chiral enhancement factors are kept in one sign in the whole range[41], as well as our structures in this paper. Besides, in our conditions, the enhancement peaks are always following the plasmon resonance peaks.

More simulations show that this kind of chiral field also exists in other shaped dimers, such as nano rod, nano rice and bowtie structures. From the above figures one can see that the chiral fields in the gap and at the ends along the dimer axis have the same handedness, while on the sides have opposite handedness. It is useful in practice because for chemically wet synthesized structures, the opposite faces or ends usually have the same crystal facet while the face and the end have different ones, thus one can selectively attach specific molecules just to the faces or ends. In SERS experiment, one can control the molecules to adsorb in gap on the hot spot. It is especially useful in single molecule ROA experiment. The huge electric field enhancement provides strong enough Raman signals and the large chiral field guarantees the chiral response of the molecule.

## Methods

All full wave numerical simulations were done by using finite element method (FEM, commercial software package, Comsol Multiphysics 5.0). The Au (Palik) objects were put on glass substrate in water environment or only in air. Non-uniform meshes were used for formatting the object accurately. The smallest mesh close to the object is 1 nm and the biggest mesh is set less than λ/6. Perfect matched layer (PML) was used to minimize the scattering from the outer boundary. The dimers were put in x-y plane. The incident light was set to 1V/m and propagates in the z direction. Electromagnetic fields on an imaginary spherical surface with a radius larger than 300 nm enclosing the structure was used to calculate the far-field scattering cross section with the Stratton−Chu formula. The absorption cross section was calculated by integrating the Ohmic heating within the Au dimer. The super chiral field was plotted with $\hat{C} = C/|C_{CPL}|$, where C is defined as $C = -\frac{\varepsilon_0 \omega}{2} \cdot Im(\vec{\boldsymbol{E}}^* \cdot \vec{\boldsymbol{B}})$, and $C_{CPL} = \pm \frac{\varepsilon_0 \omega}{2C} \cdot |\vec{\boldsymbol{E}}|^2$. The volume averaged chiral spectra were got with$\langle \hat{C} \rangle = \frac{1}{V} \int_V \ \hat{C} \cdot dV$.

## Acknowledgements

This work was supported by the National Natural Science Foundation of China (Grant Nos. 91436102, 11374353, 11474141, 11404055, 21471039 and 21473115), the Program of the Liaoning Key Laboratory of Semiconductor Light-Emitting Photocatalytic Materials.

## Author contributions

Y. Fang and M. Sun launched and supervised the project. X. Tian and Y. Fang did the analytical and numerical simulations, analyzed the data and wrote the manuscript. X. Tian, Y. Fang and M. Sun revised the manuscript.

**Competing financial interests:** The authors declare no competing financial interests.

# Supplementary Information for

# Formation of Enhanced Uniform Chiral Fields in Symmetric Dimer Nanostructures

Xiaorui Tian[1], Yurui Fang[2,*], and Mengtao Sun[3,*]

[1]Division of i-Lab, Suzhou Institute of Nano-Tech & Nano-Bionics, Chinese Academy of Sciences, Suzhou 215123, Jiangsu, China
[2]Department of Applied Physics, Chalmers University of Technology, SE-412 96, Göteborg, Sweden
[3]Beijing National Laboratory for Condensed Matter Physics, Institute of Physics, Chinese Academy of Sciences, Beijing, 100190, Beijing, China
[*]Corresponding authors: yurui.fang@chalmers.se (Y. Fang), mtsun@iphy.ac.cn (M. Sun)

## The electric dyadic Green's functions

The electric dyadic Green's tensor $\overleftrightarrow{\boldsymbol{G}_e}(\boldsymbol{r},\boldsymbol{r_0})$ is the free space field susceptibility tensor relating an electric dipole source $\boldsymbol{p_e}$ at position $\boldsymbol{r_0}$ in vacuum to the electric field $\boldsymbol{E}$ it generates at position $\boldsymbol{r}$ through[1]

$$\boldsymbol{E}(\boldsymbol{r}) = \frac{k^2}{\varepsilon_0}\overleftrightarrow{\boldsymbol{G}_e}(\boldsymbol{r},\boldsymbol{r_0})\boldsymbol{p_e} \quad (S1)$$

With

$$\overleftrightarrow{\boldsymbol{G}_e}(\boldsymbol{r},\boldsymbol{r_0}) = \frac{e^{ikr}}{r}\left[\left(\widehat{\boldsymbol{n}}\otimes\widehat{\boldsymbol{n}}-\overleftrightarrow{\boldsymbol{I}}\right)+\frac{ikr-1}{k^2r^2}\left(3\cdot\widehat{\boldsymbol{n}}\otimes\widehat{\boldsymbol{n}}-\overleftrightarrow{\boldsymbol{I}}\right)\right] \quad (S2)$$

Where $r = |\boldsymbol{r}-\boldsymbol{r_0}|$, $k = 2\pi/\lambda$ and $\widehat{\boldsymbol{n}} = \frac{\boldsymbol{r}-\boldsymbol{r_0}}{r}$.

## The coupled dipole approximation method

Let us consider many three-dimensional dipole scatters. The electric field yielded by each dipole is

$$\boldsymbol{E_j}(\boldsymbol{r}) = \frac{k^2}{\varepsilon_0}\overleftrightarrow{\boldsymbol{G}_e}(\boldsymbol{r},\boldsymbol{r_0})\boldsymbol{p_{e,j}} \quad (S3)$$

The dipole moment of each dipole can be expressed as

$$\boldsymbol{p_j} = \overleftrightarrow{\boldsymbol{\alpha_j}}\boldsymbol{E_{j,total}} = \overleftrightarrow{\boldsymbol{\alpha_j}}\left(\boldsymbol{E_{j,in}} + \sum_{k=1,k\neq j}^{N} \frac{k^2}{\varepsilon_0}\overleftrightarrow{\boldsymbol{G_e}}(\boldsymbol{r_j},\boldsymbol{r_k})\overleftrightarrow{\boldsymbol{\alpha_k}}\boldsymbol{E_{k,total}}\right) \quad (S4)$$

For two coupled dipoles, we have

$$\boldsymbol{p_{e,1}} = \overleftrightarrow{\boldsymbol{\alpha_1}}\left(\boldsymbol{E_{1,in}} + \frac{k^2}{\varepsilon_0}\overleftrightarrow{\boldsymbol{G_e}}(\boldsymbol{r_1},\boldsymbol{r_2})\boldsymbol{p_{e,2}}\right) \quad (S5)$$

$$\boldsymbol{p_{e,2}} = \overleftrightarrow{\boldsymbol{\alpha_2}}\left(\boldsymbol{E_{2,in}} + \frac{k^2}{\varepsilon_0}\overleftrightarrow{\boldsymbol{G_e}}(\boldsymbol{r_2},\boldsymbol{r_1})\boldsymbol{p_{e,1}}\right) \quad (S6)$$

We can easily get the self-consistent form of dipole moments

$$\boldsymbol{p_{e,1}} = \frac{\overleftrightarrow{\boldsymbol{\alpha_1}}\boldsymbol{E_{1,in}} + \frac{k^2}{\varepsilon_0}\overleftrightarrow{\boldsymbol{\alpha_1}}\overleftrightarrow{\boldsymbol{G_e}}(\boldsymbol{r_1},\boldsymbol{r_2})\overleftrightarrow{\boldsymbol{\alpha_2}}\boldsymbol{E_{2,in}}}{\overleftrightarrow{\boldsymbol{I}} - \frac{k^4}{\varepsilon_0^2}\overleftrightarrow{\boldsymbol{\alpha_1}}\overleftrightarrow{\boldsymbol{G_e}}(\boldsymbol{r_1},\boldsymbol{r_2})\overleftrightarrow{\boldsymbol{\alpha_2}}\overleftrightarrow{\boldsymbol{G_e}}(\boldsymbol{r_2},\boldsymbol{r_1})} \quad (S7)$$

$$\boldsymbol{p_{e,2}} = \frac{\overleftrightarrow{\boldsymbol{\alpha_2}}\boldsymbol{E_{2,in}} + \frac{k^2}{\varepsilon_0}\overleftrightarrow{\boldsymbol{\alpha_2}}\overleftrightarrow{\boldsymbol{G_e}}(\boldsymbol{r_2},\boldsymbol{r_1})\overleftrightarrow{\boldsymbol{\alpha_1}}\boldsymbol{E_{1,in}}}{\overleftrightarrow{\boldsymbol{I}} - \frac{k^4}{\varepsilon_0^2}\overleftrightarrow{\boldsymbol{\alpha_2}}\overleftrightarrow{\boldsymbol{G_e}}(\boldsymbol{r_2},\boldsymbol{r_1})\overleftrightarrow{\boldsymbol{\alpha_1}}\overleftrightarrow{\boldsymbol{G_e}}(\boldsymbol{r_1},\boldsymbol{r_2})} \quad (S8)$$

The generated total electric field at position **r** is

$$\boldsymbol{E}(\boldsymbol{r}) = \boldsymbol{E_1}(\boldsymbol{r}) + \boldsymbol{E_2}(\boldsymbol{r}) = \frac{k^2}{\varepsilon_0}\left(\overleftrightarrow{\boldsymbol{G_e}}(\boldsymbol{r},\boldsymbol{r_0})\boldsymbol{p_{e,1}} + \overleftrightarrow{\boldsymbol{G_e}}(\boldsymbol{r},\boldsymbol{r_0})\boldsymbol{p_{e,2}}\right) \quad (S9)$$

## The polarizability of the dipoles

The polarizability tensor for spherical nanoparticle is[2,3]

$$\overleftrightarrow{\boldsymbol{\alpha}}(\boldsymbol{r},\omega) = \overleftrightarrow{\boldsymbol{\alpha_0}}(\boldsymbol{r},\omega)[\overleftrightarrow{\boldsymbol{I}} - (2/3)ik_0^3\overleftrightarrow{\boldsymbol{\alpha_0}}(\boldsymbol{r},\omega)]^{-1} \quad (S10)$$

Where $\overleftrightarrow{\boldsymbol{\alpha_0}}(\boldsymbol{r},\omega)$ is the Clausius-Mossotti polarizability

$$\overleftrightarrow{\boldsymbol{\alpha_0}}(\boldsymbol{r},\omega) = \frac{3d^3}{4\pi}\left(\overleftrightarrow{\boldsymbol{\varepsilon}}(\boldsymbol{r},\omega) - \overleftrightarrow{\boldsymbol{I}}\right)\left(\overleftrightarrow{\boldsymbol{\varepsilon}}(\boldsymbol{r},\omega) + 2\overleftrightarrow{\boldsymbol{I}}\right)^{-1} \quad (S11)$$

While for Hertzian dipole, Schäferling et al.[4] used

$$\overleftrightarrow{\alpha} = \frac{\boldsymbol{E_{in}} \cdot \boldsymbol{p}}{|\boldsymbol{E_{in}}|}\overleftrightarrow{\boldsymbol{I}}\, \boldsymbol{e}^{i\frac{\pi}{2}} \quad (S12)$$

Where the pi/2 was set to indicate a plasmonic antenna induced phase delay. Actually, for real metal nanoparticles, the phase delay is gotten automatically. The dot product means only components of the electric field parallel to the diole axis can excite the dipole. While in our paper, a fixed linearly polarized light is used. The dipole coupling and incident light together decide the final dipole moments ($\overleftrightarrow{\boldsymbol{\alpha}} = p\overleftrightarrow{\boldsymbol{I}}\, \boldsymbol{e}^{i\frac{\pi}{2}}$). The simulation shows that even for the spherical metal particles, the dipole moments are more along the incident light, because the particle dipole are point dipole, so the gap is relatively big.

## The chiral field yielded by the scattering of the coupled dipoles and the incident field

The optical chirality C of the field at positon $\boldsymbol{r}$ is

$$\begin{aligned} C(\boldsymbol{r}) &= -\frac{\varepsilon_0 \omega}{2} \cdot Im\big(\boldsymbol{E}^*(\boldsymbol{r}) \cdot \boldsymbol{B}(\boldsymbol{r})\big) \\ &= -\frac{\varepsilon_0 \omega}{2} \cdot Im\big((\boldsymbol{E_{in}} + \boldsymbol{E_d})^* \cdot (\boldsymbol{B_{in}} + \boldsymbol{B_d})\big) \\ &= C_{in} - \frac{\varepsilon_0 \omega}{2} \cdot [Im(\boldsymbol{E_{in}}^* \cdot \boldsymbol{B_d}) + Im(\boldsymbol{E_d}^* \cdot \boldsymbol{B_{in}})] \quad (S13) \end{aligned}$$

Where $C_{in} = 0$ is the optical chirality of the incident linearly polarized light. Term $\boldsymbol{E_d}^* \cdot \boldsymbol{B_d} = 0$ as the electric and magnetic field vectors are orthogonal. $\boldsymbol{E_{in}}^* \cdot \boldsymbol{B_d} = 0$ because $\boldsymbol{B_d}$ is always orthogonal to the dipole $\boldsymbol{p_e}$ and thus has not parallel component with $\boldsymbol{E_{in}}$. So the chiral field yielded by the dipole is

$$C(\boldsymbol{r}) = -\frac{\varepsilon_0 \omega}{2} \cdot Im(\boldsymbol{E_d}^* \cdot \boldsymbol{B_{in}}) \quad (S14)$$

## Evaluate the gap dependent chiral fields in the gap between the two dipoles

From the chiral field expression, it is hard to see how the chiral field in the gap depends on the distance of the two dipoles. However, from the numerical simulations as well as the above two coupled dipoles theory, the two dipoles have almost the same polarization vector at the resonant wavelength. In addition, the dipole moments increase with the decrease of the gap. So we consider the moments of dipoles as $\boldsymbol{p_{e,1}} = \boldsymbol{p_{e,2}} = p(d)\widehat{\boldsymbol{p}}$, where $\widehat{\boldsymbol{p}}$ is dipole direction.

Considering this is the near field properties, the electric and magnetic dyadic Green's functions of the dipole can be approximated as

$$\boldsymbol{E}(\boldsymbol{r}) = \frac{k^2}{\varepsilon_0}\overleftrightarrow{\boldsymbol{G}_e}(\boldsymbol{r},\boldsymbol{r_0})\boldsymbol{p_e} \approx \frac{1}{4\pi\varepsilon_0}\frac{e^{ikr}}{r^3}(3\widehat{\boldsymbol{n}}(\widehat{\boldsymbol{n}}\cdot\widehat{\boldsymbol{p}})-\widehat{\boldsymbol{p}})p(d) \quad (15)$$

So the optical chirality

$$C(\boldsymbol{r}) = -\frac{\varepsilon_0\omega}{2}\cdot Im(\boldsymbol{E_d}^*(\boldsymbol{r})\cdot\boldsymbol{B_{in}}) \sim \frac{1}{r^3}(3\widehat{\boldsymbol{n}}(\widehat{\boldsymbol{n}}\cdot\widehat{\boldsymbol{p}})-\widehat{\boldsymbol{p}})\cdot p(d) \quad (16)$$

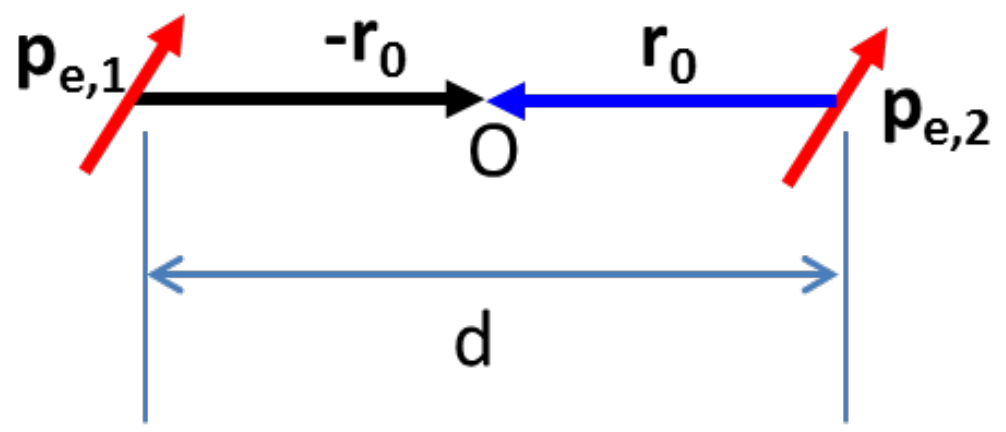

**Figure S1**. Schemetically illustrate the approximation of evaluating the gap dependent chiral field between the dipoles.

Consider the origin point O in the middle of the two dipoles, we have

$$\begin{aligned}
&\boldsymbol{E}(\boldsymbol{r}) = \boldsymbol{E_1}(\boldsymbol{r}) + \boldsymbol{E_2}(\boldsymbol{r}) \\
&\sim \frac{1}{r_1^3}(\boldsymbol{3\hat{r}_1}(\boldsymbol{\hat{r}_1}\cdot\widehat{\boldsymbol{p}})-\widehat{\boldsymbol{p}})p(d) + \frac{1}{r_2^3}(\boldsymbol{3\hat{r}_2}(\boldsymbol{\hat{r}_2}\cdot\widehat{\boldsymbol{p}})-\widehat{\boldsymbol{p}})p(d) \\
&= \frac{p(d)}{r_0^3}((\boldsymbol{3}(-\boldsymbol{\hat{r}_0})((-\boldsymbol{\hat{r}_0})\cdot\widehat{\boldsymbol{p}})-\widehat{\boldsymbol{p}}) + (\boldsymbol{3\hat{r}_0}(\boldsymbol{\hat{r}_0}\cdot\widehat{\boldsymbol{p}})-\widehat{\boldsymbol{p}})) \\
&= \frac{2p(d)}{r_0^3}(\boldsymbol{3\hat{r}_0}(\boldsymbol{\hat{r}_0}\cdot\widehat{\boldsymbol{p}})-\widehat{\boldsymbol{p}}) \qquad (17)
\end{aligned}$$

then

$$C(\boldsymbol{r}) \sim \frac{2p(d)}{r_0^3}(\boldsymbol{3\hat{r}_0}(\boldsymbol{\hat{r}_0}\cdot\widehat{\boldsymbol{p}})-\widehat{\boldsymbol{p}}) \sim \frac{p(d)}{d^3} \quad (18)$$

So as the distance of the two dipoles increases, the chiral field between the dipoles increases, following the relation $\frac{p(d)}{d^3}$ . This relation is only for chiral field of the coupled ideal dipoles, while for metal nanoparticles, it is hard to find the relation analytically. However, it is reasonable to believe that the monotonically decreasing trend is similar, and the COMSOL simulation confirmed that the chiral near field has a monotonically decreasing trend with the gap d .